\documentclass[sigconf,nonacm]{acmart}
\usepackage{xcolor}
\usepackage[utf8]{inputenc}
\usepackage{amsmath}
\usepackage{graphicx} 
\usepackage{siunitx}
\usepackage{array}
\usepackage[frozencache]{minted}
\usepackage{soulutf8}
\usepackage{tikz}
\usetikzlibrary{decorations.pathreplacing}
\usepackage{float}
\usepackage{flushend}

\usepackage{enumitem}
\usepackage{balance}
\usepackage{hyperref}
\usepackage{url}
\usepackage{xspace}
\usepackage{tcolorbox}
\usepackage{multirow}
\usetikzlibrary{shapes}
\usetikzlibrary{patterns, fit}
\usepackage{adjustbox}
\usepackage{pifont}
\setminted[java]{
xleftmargin=15pt,
framesep=2mm,
linenos=true,
breaklines=true,
tabsize=2,
encoding=utf8,
frame=none,
fontsize=\scriptsize,
escapeinside=|| 
}

\makeatletter
\newcommand{\ostar}{\mathbin{\mathpalette\make@circled\star}}
\newcommand{\make@circled}[2]{%
  \ooalign{$\m@th#1\smallbigcirc{#1}$\cr\hidewidth$\m@th#1#2$\hidewidth\cr}%
}
\newcommand{\smallbigcirc}[1]{%
  \vcenter{\hbox{\scalebox{0.97778}{$\m@th#1\bigcirc$}}}%
}

\newboolean{showcomments}
\setboolean{showcomments}{true}
\ifthenelse{\boolean{showcomments}}
 { \newcommand{\mynote}[2]{
      \fbox{\bfseries\sffamily\scriptsize#1}
        {\small$\blacktriangleright$\textsf{\emph{#2}}$\blacktriangleleft$}}}
        { \newcommand{\mynote}[2]{}}

\newcolumntype{g}{>{\columncolor{gray!30}}l}
\newcolumntype{h}{>{\columncolor{gray!30}}c}

\newcommand{\difuzer}[0]{\textsc{Difuzer}\xspace}
\newcommand{\flowdroid}[0]{\textsc{FlowDroid}\xspace}
\newcommand{\tsopen}[0]{\textsc{TSOpen}\xspace}
\newcommand{\ts}[0]{\textsc{TriggerScope}\xspace}
\newcommand{\soot}[0]{\textsc{Soot}\xspace}
\newcommand{\hm}[0]{\textsc{HsoMiner}\xspace}
\newcommand{\az}[0]{\textsc{AndroZoo}\xspace}
\newcommand{\gp}[0]{Google Play\xspace}
\newcommand{\dataset}[0]{\textsc{TriggerZoo}\xspace}
\newcommand{\jimple}[0]{\textsc{Jimple}\xspace}
\newcommand{\tool}[0]{\textsc{AndroBomb}\xspace}
\newcommand{\numberOfApps}[0]{406\xspace}
\newcommand{\numberOfAppsMalicious}[0]{240\xspace}
\newcommand{\numberOfAppsBenign}[0]{166\xspace}
\newcommand{\droidbench}[0]{\textsc{DroidBench}\xspace}
\newcommand{\androidcompass}[0]{\textsc{AndroidCompass}\xspace}
\newcommand{\andror}[0]{\textsc{AndroR2}\xspace}
\newcommand{\androct}[0]{\textsc{AndroCT}\xspace}

\begin{document}
\settopmatter{printacmref=false}
\setcopyright{none}

\title{TriggerZoo: A Dataset of Android Applications Automatically Infected with Logic Bombs} 

\author{Jordan Samhi, Tegawendé F. Bissyandé, Jacques Klein}
\affiliation{%
  \institution{SnT, University of Luxembourg, Luxembourg, \{jordan.samhi, tegawende.bissyande, jacques.klein\}@uni.lu}
  \country{}
}

\begin{abstract}
Many Android apps analyzers rely, among other techniques, on dynamic analysis to monitor their runtime behavior and detect potential security threats.
However, malicious developers use subtle, though efficient, techniques to bypass dynamic analyzers. Logic bombs are examples of popular techniques where the malicious code is triggered only under specific circumstances, challenging comprehensive dynamic analyses.
The research community has proposed various approaches and tools to detect logic bombs. Unfortunately, rigorous assessment and fair comparison of state-of-the-art techniques are impossible due to the lack of ground truth.
In this paper, we present \dataset, a new dataset of \numberOfApps Android apps containing logic bombs and benign trigger-based behavior that we release only to the research community using authenticated API.
These apps are real-world apps from \gp that have been automatically infected by our tool \tool.
The injected pieces of code implementing the logic bombs cover a large pallet of realistic logic bomb types that we have manually characterized from a set of real logic bombs.
Researchers can exploit this dataset as ground truth to assess their approaches and provide comparisons against other tools.
\end{abstract}

\maketitle

\keywords{Dataset, Android Security, Logic Bombs}

\section{Introduction}
\label{sec:introduction}

The Android operating system is the most used worldwide in mobile devices~\cite{idc_market_share}.
Hence, Android security and privacy have become one of the major concerns of researchers.
Every year, several thousands of threats are identified by antivirus companies spanning a wide range of maliciousness (e.g., trojan, adware, spyware, ransomware, etc.).
To cope with malicious code proliferation, researchers set up several approaches that rely on static analysis~\cite{7792435,doi:10.1155/2015/479174,7546513,samhi2021jucify}, dynamic analysis~\cite{10.1145/2592791.2592796,van2013dynamic,zheng2012smartdroid}, machine-learning based analysis~\cite{6298824,6735264,10.1007/978-3-030-87839-9_4}, or hybrid approaches~\cite{10.1007/978-3-319-56991-8_51,8441295,brumley2008automatically}.

Nowadays, malicious developers build their codebase to avoid detection from analyzers~\cite{fratantonio2016triggerscope, samhi2021difuzer,10.1007/978-3-030-49785-9_11,9524530,10.1007/978-3-030-01701-9_10}.
A notable technique used to bypass dynamic analyses consists in employing \emph{logic bombs} that allow the malicious code to be triggered only under specific circumstances (e.g., at a specific date). In recent years, researchers have therefore proposed various techniques to uncover logic bombs in Android applications (apps)~\cite{fratantonio2016triggerscope, samhi2021difuzer,pan2017dark}.
However, a common challenge in advancing state of the art is the lack of shared benchmarks for the assessment and fair comparison of literature approaches. 

The research literature already proposed various datasets of Android apps to encourage reproducibility and comparison between different approaches. For instance, Allix et al. proposed Androzoo~\cite{Allix:2016:ACM:2901739.2903508}, a growing repository now including about 18 Million Android apps.
Arzt et al. released \droidbench~\cite{arzt2014flowdroid}, a test suite to evaluate Android taint analysers.
Nielebock et al. proposed \androidcompass~\cite{9463151} as a dataset of Android compatibility checks.
Recently, Wendland et al.~\cite{9463119} released \andror, a dataset of bug reports related to Android apps and Li et al.~\cite{9463081} released \androct, a large-scale dataset of runtime traces of benign and malicious Android apps. However, in the research directions related to logic bombs, the community faces a challenge to build a comprehensive dataset due to the known difficulties in detecting logic bombs. Indeed,  even if an app is detected as malware, identifying a logic bomb in malware requires extensive manual inspection and strong expertise. Logic bombs are often simple \emph{if statements} with "unusual" conditional expressions. Yet, it is far from being trivial to distinguish a "logic bomb condition" from a "legitimate and normal condition".
The research community lacks an important artifact in the logic bomb detection domain, i.e., an Android app dataset that contains logic bombs with information about their localization in the apps.

In this work, we propose a new dataset of Android apps containing logic bombs and benign trigger-based behavior to the research community.
This dataset, named \dataset, contains \numberOfApps apps, from which \numberOfAppsMalicious are infected with logic bombs, and \numberOfAppsBenign apps contain benign trigger-based behavior.
It was generated by applying our dedicated tool, \tool, on 2000 apps from \gp. 
\dataset is meant to facilitate research on logic bomb detection.
Specifically, \dataset will serve as a base for new approaches to detect logic bombs to assess new tools and compare with other approaches.
Besides, since \tool has been developed with a modular approach, it is easy to add new trigger and behavior types.

The main contributions of our work are as follows:
\begin{itemize}[leftmargin=*]
\setlength\itemsep{0em}
    \item We propose \dataset, a new reusable dataset of \numberOfApps Android apps infected with trigger-based behavior, and their localization with 10 trigger types and 14 behavior types.
    \item We also propose \tool, an extensible framework to inject trigger-based behaviors into Android apps automatically. 
    \item We provide performance results of two state-of-the-art works \difuzer and \ts on the \dataset dataset.
\end{itemize}

\dataset apps are made available in the AndroZoo repository, where they are responsibly shared with authenticated researchers only.
\dataset apps' hashes, and labels are available at: \\
\centerline{\url{https://github.com/JordanSamhi/TriggerZoo}\footnote{DOI: 10.5281/zenodo.5907916, Access: \url{https://doi.org/10.5281/zenodo.5907916}}}

In the same way, and to avoid encouraging malware development, \tool is only available to authenticated researchers.
\tool's instructions and appropriate files are available at: \\
\centerline{\url{https://github.com/JordanSamhi/AndroBomb}\footnote{DOI: 10.5281/zenodo.5907924, Access: \url{https://doi.org/10.5281/zenodo.5907924}}}
\section{Logic bombs}
\label{sec:background}

A logic bomb is a piece of malicious code triggered under very specific circumstances.
It means that the malicious code is segregated from the normal execution of the benign code and is only triggered under specific criteria.
For instance, if attackers want to target devices in Russia \emph{and} that have two cameras (i.e., a back and a front one), they can rely on the code illustrated in Listing~\ref{code:motivation}.

\begin{listing}[h]
\begin{minted}{java}
public class MainActivity extends Activity {
  protected void onCreate(Bundle b) {
    TelephonyManager tm = (TelephonyManager) this.getSystemService("phone");
    String countryCode = tm.getNetworkCountryIso();
    if(countryCode.equals("RU")){
      Camera.PictureCallback cb = new Camera.PictureCallback(){
        public void onPictureTaken(final byte[] b, Camera c)
          {/* send photo to http server + data theft*/}
      }
      if (Camera.getNumberOfCameras() >= 2)
        {Camera.open(1).takePicture(null, null, cb);}
    }[...] // benign behavior 
  }
}
\end{minted}
    \caption{An illustration of a logic bomb in an Android app}
    \label{code:motivation}
\end{listing}

Indeed, on line 5, the developer verifies if the device is connected to a Russian mobile network (the ISO-3166-1 alpha-2 country code RU represents the Russian country).
On lines 11-13, a photograph is taken from the front camera if the device has at least two cameras and is sent over the network, breaking the user privacy.
In this case, the first specific circumstance is that the device is connected to a Russian mobile network (i.e., the Russian population is targeted), and the second one is that the device contains two cameras (i.e., eliminating emulators and old devices.).
This example shows that the malicious code is only executed on specific conditions that are unlikely to happen when dynamically analyzing the app in a sandbox. The app's maliciousness will therefore escape detection.

In the rest of the paper, 
we refer to \emph{trigger type} for the type of the condition that triggers the logic bomb or the benign trigger-based behavior (e.g., at a specific time), 
and to \emph{guarded code type} for the type of the behavior triggered (e.g., data theft).
The interested reader can refer to~\cite{samhi2021difuzer} for further details on the logic bomb concept.
\section{Dataset Construction Methodology}
\label{sec:dataset}

In this section, we first describe \tool used to construct \dataset.
Secondly, we give details about the process we followed to generate our dataset and describe it.

\subsection{\tool: Automatically Infect Apps}

\begin{figure*}[ht]
    \centering
    \begin{adjustbox}{width=.8\linewidth,center}
    \input{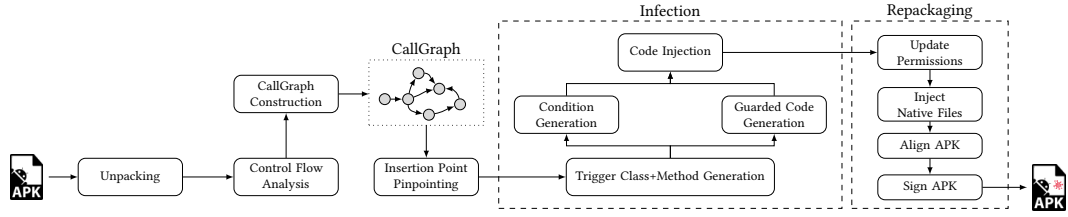}
    \end{adjustbox}
    \caption{Overview of the \tool approach to infect an Android app.}
    \label{fig:overview}
\end{figure*}

\tool has been designed to inject a trigger-based behavior (malicious or benign) in a specific location in an Android app. This trigger-based behavior is characterized by a \emph{trigger type} (e.g., time check) and a \emph{guarded code type} (e.g., the stealing of private information).
In Figure~\ref{fig:overview}, we present an overview of the \tool approach.
\tool is made of three main parts:
 \ding{182} a mechanism for pinpointing an insertion point based on callgraph construction and control flow analysis that serves to identify a method in which a trigger-based behavior can be inserted;
 \ding{183} an infection step where a trigger-based behavior is generated given a \emph{condition type} and a \emph{guarded code type}, and inserted in the insertion point;
 and \ding{184} a repackaging step where the APK is updated with new permissions (if required), new native code files (if required), aligned and signed to generate an infected APK file.

\subsubsection{Insertion point Pinpointing Mechanism}
We intend to inject trigger-based behavior into Android apps.
The idea is to inject these trigger-based behavior in methods \ding{182} highly likely to be executed at runtime, \ding{183} present in the developer code.
To find these methods, \tool relies on \flowdroid~\cite{arzt2014flowdroid} and \soot~\cite{vallee2010soot} which provide a control flow analysis that is used to generate a callgraph.
We set \flowdroid to use the SPARK algorithm~\cite{lhotak2003spark} to generate a callgraph.
Then, \tool builds a set of methods $M$ that contains all the methods in the APK that are declared in a class for which the fully qualified name starts with the app's package name, i.e., it is a developer class. Indeed, we want to inject the logic bomb in the developer code to simulate malicious intent from the developer.
$M$ is then filtered to produce a new set of methods $M_{cg}$ that only retains methods that are present in the callgraph previously generated, i.e., they may be called during execution.
Eventually, \tool randomly choses a method in $M_{cg}$ that will act as the \emph{insertion point}.

\subsubsection{Infection}

To infect an app, \tool needs two information:
\ding{182} the trigger type used to activate the trigger-based behavior;
and \ding{183} the guarded code type.
This information is not static and is given as options to \tool.
As already mentioned, \tool relies on \flowdroid, hence \tool manipulates \jimple code~\cite{vallee1998jimple} which is the language used to perform code instrumentation~\cite{instrumenting} and code injection.
After dynamically generating the class and the method in which the code generated will lie, \tool generates the trigger (according to the given type) and the guarded code (according to the given type).
These pieces of code are merged to constitute a single entity (i.e., condition + code triggered) and injected into the insertion point.
Note that \emph{trigger types} and \emph{guarded code types} have been chosen from existing logic bombs found in previous research and reverse-engineering, as well as benign trigger-based behavior found in the same way.

\subsubsection{Packaging}

After infecting the app's code, one has to take care of any collateral effect.
Therefore, \tool adds any permission needed for the code injected~\cite{10.1145/2382196.2382222}.
For instance, if \tool injects a piece of code that steals the current location of the device and sends it over HTTP, \tool also injects the following permissions to the AndroidManifest.xml file:
\begin{itemize}
    \item android.permission.ACCESS\_COARSE\_LOCATION
    \item android.permission.ACCESS\_FINE\_LOCATION
    \item android.permission.INTERNET
\end{itemize}
This is to ensure that no error occurs at execution time.

Besides, \tool can inject pieces of code that might invoke native code~\cite{ndk}.
However, to invoke a native function, the APK should contain the related .so file (i.e., native libraries).
Thus, \tool also injects the adequate .so files needed to invoke a native library.
The authors of this paper have developed these libraries, their source code is available in the project's repository.

Eventually, the resulting APK is aligned~\cite{zipalign}, and signed~\cite{signing}.
Therefore, it can be installed on any device or emulator.

As \tool only injects pieces of code that do not change the state of the initial app, the overall behavior remains unchanged.
The infected app can be dynamically analyzed and monitored using emulators or statically analyzed to search for \emph{potential} logic bombs.
In addition, \tool has been developed in a modular manner which allows the community to easily add new \emph{trigger types} and new \emph{guarded code types} to generate new apps for this dataset which is by design prone to evolve.
We believe that this work will serve the community and advance the logic bomb detection research field.

\subsection{\dataset}

To generate \dataset, we relied on \tool which can, at the time of writing, handle the trigger and guarded code types described in Tables~\ref{table:trigger_types} and~\ref{table:guarded_code_types}.
The authors have carefully chosen the trigger and guarded code types from existing logic bomb and benign trigger-based behavior that have been reverse-engineered. 
In addition, the literature~\cite{fratantonio2016triggerscope,samhi2021difuzer,9524530} describes several use-cases for which the authors extracted the trigger and guarded code types.
Even if we cannot guarantee to cover all possible types of trigger and guarded code types, by relying on these state-of-the-art works, we are confident that \dataset covers a large proportion of logic bomb types that could exist in the wild.

\subsubsection{Dataset Construction}

We randomly collected 2000 \gp apps from the \az dataset as our initial set.
Then, for each app, we applied \tool with the trigger and guarded coded types randomly generated among those available in Tables~\ref{table:trigger_types} and~\ref{table:guarded_code_types}. Each app is instrumented to receive one single trigger-based behavior.
The resulting dataset, namely \dataset, comprises \numberOfApps Android apps infected with trigger-based behavior.
There are several reasons why \tool was not able to generate an infected app for all of the 2000 apps:
\begin{itemize}[noitemsep,topsep=0pt,leftmargin=*]
    \item No insertion point was found in the app due to our strong constraint: we only consider methods that are in classes for which the fully qualified name starts with the app package name. (\num{28.9}\%).
    \item The infected APK could not be repackaged due to some limitations of third-party software. For instance, (1) \soot could not handle multi dex APKs for apps using an Android API level lower than 22, or (2) The ManifestEditor library crashes due to buffer underflow (\num{54.5}\%).
    \item \tool crashes since for some apps the methods added during infection do not exist yet because they were added in a subsequent Android API level (\num{16.6}\%).
    % (e.g., getImei() was added in Android API level 26. Therefore it cannot be found in previous releases) (\num{16.6}\%).
\end{itemize}

However, these \tool limitations are not critical since its final goal is not to be 100\% operational for a specific task (e.g., malware detection~\cite{10.1145/3029806.3029823} and GDPR compliance~\cite{9251060}) but to construct a valuable dataset for the community, which it was able to achieve.

\begin{table}[ht]
    \centering
    \caption{Trigger types handled by \tool to generate \dataset}
    \begin{adjustbox}{width=.8\columnwidth,center}
        \begin{tabular}{|l|l|l|}
            \hline
            \multicolumn{2}{|c|}{\textbf{Trigger Types}} \\ \hline
            \textbf{Type} & \textbf{Description} \\ \hline
            time & at a specific time or date \\ \hline
            location & at a specific location \\ \hline
            sms & if a specific sms is received \\ \hline
            network & if Wi-Fi available or specific http response received \\ \hline
            build & if specific Build.MODEL/PRODUCT/FINGERPRINT are set \\ \hline
            camera & if the device possesses cameras \\ \hline
            addition & a dummy test with a simple addition \\ \hline
            music & if some music is active \\ \hline
            is\_screen\_on & if device in interactive state \\ \hline
            is\_screen\_off & if device not in interactive state \\ \hline
        \end{tabular}
    \end{adjustbox}
    \label{table:trigger_types}
\end{table}

\begin{table}[ht]
    \centering
    \caption{Guarded Code types handled by \tool to generate \dataset}
    {\begin{center}$\bigcirc$ = benign behavior,
    $\ostar$ = malicious behavior\end{center}}
    \begin{adjustbox}{width=\columnwidth,center}
        \begin{tabular}{|l|l|l|c|}
            \hline
            \multicolumn{3}{|c|}{\textbf{Guarded Code Types}} \\ \hline
            \textbf{Type} & \multicolumn{2}{|c|}{\textbf{Description}} \\ \hline
            return & no behavior & $\bigcirc$ \\ \hline
            sms\_imei & send the device imei number by sms & $\ostar$\\ \hline
            stop\_wifi & deactivate the device Wi-Fi connection & $\ostar$\\ \hline
            write\_string & write a constant to a file in the device's memory & $\bigcirc$ \\ \hline
            write\_phone\_number & write the phone number to a file in the device's memory & $\ostar$\\ \hline
            set\_text & set a constant to be dispayed on the screen & $\bigcirc$ \\ \hline
            sms\_string & send a constant by sms & $\bigcirc$ \\ \hline
            http\_location & sends the current location to remote server using http & $\ostar$\\ \hline
            set\_text\_reflection & set a constant to be dispayed on the screen using reflection & $\bigcirc$ \\ \hline
            exit & exits the app & $\ostar$\\ \hline
            native\_log\_string & log a constant using native code & $\bigcirc$ \\ \hline
            native\_log\_model & log the Build.MODEL information using native code & $\ostar$\\ \hline
            native\_write\_phone\_number & writes the phone number to a file using native code&  $\ostar$\\ \hline
            native\_phone\_number\_network & sends the phone number to a remote server using native code & $\ostar$\\ \hline
        \end{tabular}
    \end{adjustbox}
    \label{table:guarded_code_types}
\end{table}

\subsubsection{Dataset Description}

\dataset is composed of several files referenced in the project's repository:

\begin{itemize}[noitemsep,topsep=0pt,leftmargin=*]
    \item original\_apps: SHA256 hashes of the \numberOfApps original apps collected from \az.
    \item infected\_apps: SHA256 hashes of the \numberOfApps infected apps.
    \item original\_to\_infected\_correspondence: links the original and infected apps.
    \item triggerzoo\_labeled\_dataset: the labeled dataset.
\end{itemize}

\dataset is only available to authenticated researchers to have an access to \az.
Indeed, \az offers the authentication proxy for serving only the research community.

\textbf{Format of the labeled dataset}.
The \emph{triggerzoo\_labeled\_dataset} file in the project's repository describes in detail \dataset with the fields available in Table~\ref{table:fields}.
Each line of this file is composed of these 7 fields describing an app that has been infected.

\begin{table}[ht]
    \centering
    \caption{\dataset fields}
    \begin{adjustbox}{width=0.9\columnwidth,center}
        \begin{tabular}{|l|l|l|}
            \hline
            \multicolumn{2}{|c|}{\textbf{Field}} \\ \hline
            \textbf{Type} & \textbf{Description} \\ \hline
            sha256\_original\_app & The sha256 hash of the original app\\ \hline
            class\_infected & The class infected \\ \hline
            component\_type & The component type of the class infected \\ \hline
            method\_infected & The method infected \\ \hline
            trigger\_type & The trigger type used to infect the app \\ \hline
            guarded\_code\_type & The guarded code type used to infect the app \\ \hline
            depths & Depths of the insertion point method in the app callgraph \\ \hline
        \end{tabular}
    \end{adjustbox}
    \label{table:fields}
\end{table}

\textbf{Trigger and behavior types}.
\dataset's repository shows two plots illustrating the number of apps infected with specific trigger and guarded code types.
We can see that \dataset covers a larger panel of trigger and guarded code types, and their combination.
Note that \dataset covers 137 unique combinations of trigger and guarded code types.
Besides, based on the guarded code types, \dataset comprises 240 apps with malicious trigger-based behavior, and 166 with benign trigger-based behavior.

\textbf{Apps categories}.
\dataset's project repository shows a third plot illustrating the different categories of \dataset's apps.
We were able to retrieve the category of 263 apps from the \numberOfApps in \dataset.
This is because the remaining 143 apps were not available on \gp (i.e., they were removed from \gp after being crawled by \az).
We can see that \dataset covers a large panel of categories, i.e., 34 unique categories.

\begin{figure}[h]
    \centering
    \includegraphics[width=.9\linewidth]{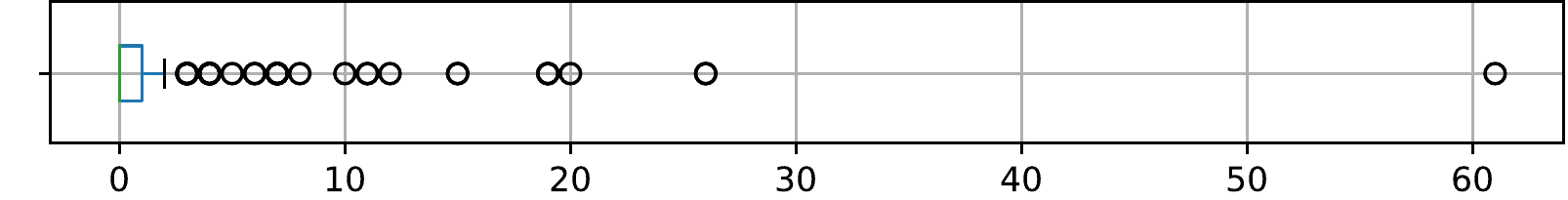}
    \caption{Infected methods' callgraph depths in \dataset}
    \label{fig:depths}
\end{figure}

\textbf{Component types and Insertion Point Depth}.
We remind that \dataset is built with \tool that injects logic bombs or benign trigger-based behavior at random locations in the callgraph of the app developer code. With this process, among our 406 apps, the trigger-based behavior have been inserted in methods of the \texttt{Activity} components for 381 apps, and \texttt{Service} components for 25 apps.
Figure~\ref{fig:depths} shows that for most apps, the callgraph depth of the insertion point is low.

\vspace{-.4cm}

\subsubsection{Dataset Validation}
To ensure that the apps are infected by \tool, we randomly selected a statically significant sample of 118 apps from the \numberOfApps supposedly infected apps, with a confidence level of 99\% and a confidence interval of $\pm$ 10\%, to manually inspect them.
We confirm that 100\% of the apps manually analyzed are infected, i.e., they contain the newly added code, the AndroidManifest.xml file is updated with new permissions (if needed in function of the APIs injected), and native files are indeed present in the app (if required in function of the APIs injected).
Also, to ensure that the apps are not faulty and can be dynamically analyzed, we tried to install and run them on emulators.
We confirm that 100\% of the 118 apps manually analyzed can be installed and run without any problem.
Hence, the instrumentation and repackaging processes do not impact the installation or the runtime processes.

\section{Importance of \dataset}
\label{sec:related_work}

Several state-of-the-art approaches have been proposed to detect logic bombs. 
\ts was proposed in~\cite{fratantonio2016triggerscope} as an automated tool to detect logic bombs.
It relies on static analysis techniques such as: symbolic execution, control flow analysis, predicate recovery, and control dependency. Recently, Samhi et al.~\cite{9524530} have released an open-source version of \ts that they named \tsopen. 
Pan et al.~\cite{pan2017dark} proposed \hm, an approach relying on static analysis techniques such as: control flow analysis, backward dependency graph, and trigger analysis.
In addition, the authors proposed a machine-learning approach to automatically sort \emph{hidden sensitive operations} (HSO).
Recently, Samhi et al.~\cite{samhi2021difuzer} proposed \difuzer, an approach to triage logic bombs among \emph{suspicious hidden sensitive operations} (SHSO).
\difuzer relies on instrumentation techniques, taint tracking to identify SHSO entry points and triggers an anomaly detection engine to detect abnormal triggers.

The common points for these approaches are: \ding{182} they do not assess their tool on existing benchmarks. Hence they cannot measure standard precision, recall, and f-1 score measures; \ding{183} they cannot compare their respective approach against each other tools.
To cope with these limitations, \dataset can be used to measure existing and future tools performances and will allow fair comparison.

In a first attempt to compare state of the art approaches, and to  assess \dataset's usefulness, we executed  \difuzer~\cite{samhi2021difuzer} and \tsopen~\cite{9524530} on the \numberOfApps apps present in \dataset to search for logic bombs.
Results are available in Table~\ref{table:difuzer}.
We note that in the original publication of \difuzer, precision and recall metrics were provided based on a-posteriori manual checking, given the lack of benchmark. Thanks to \dataset, comparisons such as the one presented in Table~\ref{table:difuzer} will be readily possible. Furthermore, note that the yielded results, with recall at 58\%, suggest that \dataset is "difficult" and will contribute to challenge future approaches. 
As a last remark, the a-posteriori comparison results presented in the \difuzer's paper~\cite{samhi2021difuzer} are confirmed on the \dataset dataset. 
Indeed, as shown in Table~\ref{table:difuzer}, \difuzer clearly outperforms \ts.

\begin{table}[ht]
    \centering
    \caption{\difuzer \& \tsopen results on \dataset}
    \vspace{-.3cm}
    \begin{adjustbox}{width=0.9\columnwidth,center}
        \begin{tabular}{|l|l|l|l|l|l|}
            \hline
            &  \multirow{2}{*}{\textbf{\# apps}}& \multicolumn{2}{c|}{\textbf{\difuzer}} &\multicolumn{2}{c|}{\textbf{\tsopen}} \\ 
            %\hline
            \cline{3-6}
            &  & \textbf{\# analyzed} & \textbf{\# flagged}&  \textbf{\# analyzed} & \textbf{\# flagged }\\ 
            \hline
            Malicious triggers & 240 & 230 & 134 &  215 & 32 \\
            \hline
            Benign triggers    & 166 & 156 & 41  & 148 & 15 \\ 
            \hline
            \multicolumn{2}{|l|}{\textbf{Precision}} & \multicolumn{2}{c|}{76.6\%} &\multicolumn{2}{c|}{68.1\%} \\
            \hline
            \multicolumn{2}{|l|}{\textbf{Recall}} &  \multicolumn{2}{c|}{58.3\%} &\multicolumn{2}{c|}{14.9\%} \\
            \hline
           \multicolumn{2}{|l|}{\textbf{F$_1$ score}}  &  \multicolumn{2}{c|}{66.2\%} &\multicolumn{2}{c|}{24.4\%} \\
            \hline
        \end{tabular}
    \end{adjustbox}
    \label{table:difuzer}
\end{table}
\vspace{-.5cm}
\section{Conclusion}
\label{sec:conclusion}

In this paper, we presented two artifacts: \ding{182} \dataset: a new evolving dataset of Android apps containing logic bombs and benign trigger-based behavior; \ding{183} \tool: a new framework to infect Android apps with logic bombs and benign trigger-based behavior.
\dataset is meant to facilitate future logic bomb detectors assessment and comparison.
We also provide results of two state-of-the-art approaches, i.e., \difuzer and \ts, on \dataset, and confirm previous literature results.
We believe that the research community will rely on this dataset to propose new approaches to detect logic bombs, thus improving Android apps security and privacy.
\dataset is not frozen. It can evolve using \tool to generate new samples with new logic bomb schemes.
Both \dataset and \tool serve the community and support the logic bomb detection research direction.
\section{Acknowledgment}
\label{sec:acknowledgment}
This work was partly supported 
(1) by the Luxembourg National Research Fund (FNR), under projects Reprocess C21/IS/16344458 the AFR grant 14596679, 
(2) by the SPARTA project, which has received funding from the European Union's Horizon 2020 research and innovation program under grant agreement No 830892, 
(3) by the Luxembourg Ministry of Foreign and European Affairs through their Digital4Development (D4D) portfolio under project LuxWAyS,
and (4) by the INTER Mobility project Sleepless@Seattle No 13999722.

\newpage

\bibliographystyle{ACM-Reference-Format}
\bibliography{bib}

\end{document}